\documentclass[twocolumn,showpacs,preprintnumbers]{revtex4}
\usepackage{graphicx}
\usepackage{dcolumn}
\usepackage{bm}
\begin{document}
\title{Azimuthal electric field in a static rotationally symmetric
(2+1)-dimensional spacetime}
\author{Mauricio Cataldo}
\altaffiliation{mcataldo@ubiobio.cl}
\affiliation{Departamento de F\'\i sica, Facultad de Ciencias,
Universidad del B\'\i o-B\'\i o, Avenida Collao 1202, Casilla 5-C,
Concepci\'on, Chile.\\}
\date{\today}

\begin{abstract}
The fundamental metrics, which describe any static
three-dimensional Einstein-Maxwell spacetime (depending only on a
unique spacelike coordinate), are found. In this case there are
only three independent components of the electromagnetic field:
two for the vector electric field and one for the scalar magnetic
field. It is shown that we can not have any superposition of these
components of the electric and magnetic fields in this kind of
static gravitational field. One of the electrostatic
Einstein-Maxwell solutions is related to the magnetostatic
solution by a duality mapping, while the second electrostatic
gravitational field must be solved separately. Solutions induced
by the more general (2+1)-Maxwell tensor on the static
cylindrically symmetric spacetimes are studied and it is shown
that all of them are also connected by duality mappings.

\pacs{04.20.Jb}
\end{abstract}
\maketitle \preprint{APS/123-QED}

In (3+1)-gravity coupled to Maxwell electrodynamics, the
Reissner-Nordstr\"{o}m black hole solution describes a vacuum
static spherically symmetric spacetime endowed with electric
$q_{e}$ or magnetic $q_{m}$ charge. If one includes the
cosmological constant $\Lambda $, the
Reissner-Nordstr\"{o}m-Kottler solution may be written
as~\cite{Kottler}
\begin{equation}
ds^{2}=-f(r)dt^{2}+\frac{dr^{2}}{f(r)}+r^{2}(d\theta ^{2}+\sin
^{2}\theta d\phi ^{2})
\end{equation}
where
\begin{eqnarray}
f(r)=1-\frac{2MG}{r}+\frac{4\pi
G(q_{e}^{2}+q_{m}^{2})}{r^{2}}-\frac{1}{3} \Lambda \,r^{2}.
\end{eqnarray}
Here, the solution has radial electric and magnetic fields given by $%
E(r)=q_{e}/r^{2}$ and $B(r)=q_{m}/r^{2}$ respectively. The case
$q_{m}=0$ describes a rest charge at the origin of spherical
coordinates and $q_{e}=0$ may be interpreted as a magnetic
monopole. The magnetically charged Reissner-Nordstr\"{o}m solution
is connected to the electrically charged solution, since all the
cases with an electric field without sources can be reformulated
to be the corresponding cases with a magnetic field (dual to the
initial electric one) or mixtures of both fields (through a
duality rotation). In all these cases the form of the metric is
the same.

The Einstein-Maxwell theories in (2+1)-dimensions without
cosmological constant have been discussed by several
authors~\cite{Gott1,Giddings,Gott}. In~\cite{Gott} the authors
found the first three-dimensional static electrically charged
space-time. Later the inclusion of the cosmological constant was
considered. In (2+1)-dimensions there exist the electric and
magnetic analogs to the Reissner-Nordstr\"{o}m-Kottler solution.
They are the static Ba\~{n}ados-Teitelboim-Zanelli (BTZ) black
hole ($E=q_{e}/r$)~\cite{BTZ}
\begin{eqnarray}
\label{BTZ} ds^{2} &=&(r^{2}/l^{2}-q_{_{e}}^{2}ln\,r-M)dt^{2}
\\
&&-\frac{dr^{2}}{(r^{2}/l^{2}-q_{_{e}}^{2}ln\,r-M)}-r^{2}d{\theta}^{2},
\nonumber
\end{eqnarray}
and the Hirschmann-Welch (HW) magnetic solution
($B=q_{m}/r$)~\cite{HW,Cataldo}
\begin{eqnarray}
\label{HWS} ds^{2} =(r^{2}/l^{2}-M)dt^{2}
 -\frac{r^{2}dr^{2}}{\tilde{g}(r)(r^{2}/l^{2}-M)}
- \tilde{g}(r)d\theta^{2},
\end{eqnarray}
where $\tilde{g}(r)=r^{2}+q_{_{m}}^{2}ln(r^{2}/l^{2}-M)$, $\Lambda
=-1/l^{2}<~0$, $q_{_{e}}$ is the electric charge, $q_{_{m}}$
magnetic charge and $M$ is the mass of the BTZ black hole. Both
solutions are asymptotically anti-de Sitter, like the
Reissner-Nordstr\"{o}m-Kottler solution for $\Lambda <0.$ In
Ref.~\cite{Cataldo} it is shown that the both
solutions~(\ref{BTZ}) and~(\ref{HWS}) are connected by a duality
transformation. Rotating (2+1)-Einstein-Maxwell solutions have
been obtained by several authors using an appropriate coordinate
transformation~\cite{Clement,Teitelboim1}. Others solutions also
have been obtained assuming self-dual (or anti-self-dual)
condition imposed on the orthonormal basis components of the
electric and magnetic fields~\cite{Kamata,Kamata1,Cataldo1}. Note
that the presence of divergences at spatial infinity in the mass
and angular momentum is an usual feature in electrically charged
solutions~\cite{Chan}. The authors of the Ref.~\cite{Kamata1}
and~\cite{Clement1,Fernando} regularize these divergences by
taking into account a boundary contribution or by the introduction
of a topological Chern-Simons term respectively. We note also that
the magnetic solution~(\ref{HWS}) has been extended in order to
include (i) angular momentum, (ii) the definition of conserved
quantities, (iii) upper bounds for the conserved quantities
themselves, and (iv) a new interpretation for the magnetic field
source~\cite{Lemos1}.

In this letter we shall show that the above description for he
static case is incomplete, and then we will focus our attention on
a complete description for all electrostatic and magnetostatic
spacetimes in (2+1)-dimensions and on the Einstein-Maxwell theory
for cylindrically static spacetimes.

In (3+1)-dimensions the Maxwell tensor and its dual are two-forms,
but in (2+1)-dimensions the Maxwell tensor is a two-form and its
dual is a one-form *$F_{a}=1/2\epsilon _{abc}F^{bc}$
($a,b,c=0,1,2$). This means that the duality transformation which
takes place for the Reissner-Nordstr\"{o}m-Kottler black hole does
not take place for (2+1)-dimensions.

Now we shall study the three-dimensional Einstein-Maxwell theory
for any static spacetime which depends only on a unique spacelike
coordinate. The fundamental metric which describes any such static
(2+1)-dimensional spacetime manifold may be written (in the
coordinate system which employs the Killing parameter, t, as one
of the coordinates) as~\cite{Wald}
\begin{eqnarray}
\label{metricageneral} ds^2=e^{2 \alpha(x^1)} dt^2- \sum^2_{a,b=1}
h_{ab}(x^1)dx^{a}dx^b,
\end{eqnarray}
where $e^{2 \alpha(x^1)}=\xi^{c} \xi_c$ ($c=0,1,2$), being
$\xi^{c}$ a timelike Killing vector field. By an appropriate
change of coordinate the orthogonal part to $\xi^c$ may be
converted into a diagonal form (see for
example~\cite{Chandrasekhar}) and then the
metric~(\ref{metricageneral}) may be written in the following form
($t=x^0$)
\begin{eqnarray}
\label{metrica estatica} ds^2=e^{2 \alpha} \, (dx^0)^2-e^{2 \beta}
\, (dx^1)^2-e^{2 \gamma} \, (dx^2)^2,
\end{eqnarray}
where $\alpha, \beta$ and $\gamma$ are all functions of the
spacelike coordinate $x^1$.

In (2+1)-dimensions the Maxwell tensor has the form
\begin{eqnarray}
\label{siete} F=-E_1 dx^0\wedge dx^1- E_{2} dx^0\wedge dx^2
\nonumber \\ +B dx^1 \wedge dx^2,
\end{eqnarray}
i.e. the Maxwell tensor has only three independent components: two
for the vector electric field and one for the scalar magnetic
field.

We now introduce an orthonormal 1-form basis given by
\begin{eqnarray}
\label{tetrada} \theta^{(0)}=e^{\alpha(x^1)} dx^0,
\theta^{(1)}=e^{\beta(x^1)} dx^1, \theta^{(2)}=e^{\gamma(x^1)}
dx^2.
\end{eqnarray}
The components of the electric and magnetic field measured by a
locally lorentzian observer in this orthonormal basis are
$E_{(0)}$, $E_{(1)}$ and $B_{()}$. Then, the (2+1)-Maxwell
tensor~(\ref{siete}) may be written as
\begin{eqnarray}
\label{2+1 tensor de Maxwell} F=E_{(1)}(x^1) \, \theta^{(1)}\wedge
\theta^{(0)}+E_{(2)}(x^1) \, \theta^{(2)}\wedge \theta^{(0)}
\nonumber \\ +B_{()}(x^1) \theta^{(1)} \,  \wedge \theta^{(2)}.
\end{eqnarray}
The stress tensor of the electromagnetic field we shall take in
the same form as in (3+1)-dimensions, i.e.
\begin{eqnarray}
\label{tensor electromagnetico}
    T_{(a)(b)} \, \theta^{(a)} \, \otimes \, \theta^{(b)} =
 - \frac{1}{4\pi}\, \left[ F_{(a)(c)} \, F_{(b)}^{ \,\,\,\,\,\,
 (c)}\right.
\nonumber \\ \left. -\frac{1}{4} \, g_{(a)(b)} \, F_{(c)(d)} \,
F^{(c)(d)} \right] \, \theta^{(a)} \, \otimes \, \theta^{(b)},
\end{eqnarray}
since there is an ambiguity in the definition of the gravitational
constant $G$ (there is no Newtonian gravitational limit in
(2+1)-dimensions).

In the basis~(\ref{tetrada}) the components of the electromagnetic
energy-momentum tensor are
\begin{eqnarray}
T^{2+1}= -\frac{1}{8 \pi}
\left[-(E^2_{(1)}+E^2_{(2)}+B^2_{()})\theta^{(0)}\otimes
\theta^{(0)} \right. \nonumber \\
+(E^2_{(1)}-E^2_{(2)}-B^2_{()})\theta^{(1)}\otimes \theta^{(1)}
\nonumber \\ -(E^2_{(1)}-E^2_{(2)}+B^2_{()})\theta^{(2)}\otimes
\theta^{(2)} \\ \nonumber
 +2 E_{(2)} B_{()}(\theta^{(0)}\otimes \theta^{(1)}+\theta^{(1)}
\otimes \theta^{(0)})
\\
\nonumber -2 E_{(1)} B_{()}(\theta^{(0)}\otimes
\theta^{(2)}+\theta^{(2)}\otimes \theta^{(0)})
\\ \nonumber
\left. +2 E_{(1)} E_{(2)}(\theta^{(1)}\otimes
\theta^{(2)}+\theta^{(2)}\otimes \theta^{(1)})
 \right].
\end{eqnarray}
Then we have the following non-trivial components for Einstein's
equations with the cosmological constant:
\begin{eqnarray}
\label{cero-cero} e^{- 2 \beta}\,\left(\gamma\, ''+ \gamma\, '
\,^{2} - \gamma\, ' \, \beta\, ' \right) =  \nonumber \\ - \Lambda
- \frac{\kappa}{8 \pi} \, (E^2_{(1)}+E^2_{(2)}+B^2_{()}),
\end{eqnarray}
\begin{eqnarray}
\label{uno-uno} - \alpha\, ' \, \gamma\, ' \,e^{- 2\, \beta}=
\Lambda +  \frac{\kappa}{8 \pi} (E^2_{(1)}-E^2_{(2)}-B^2_{()}),
\end{eqnarray}
\begin{eqnarray}
\label{dos-dos} e^{- 2 \beta}\,\left(  \alpha\, ' \, \beta\, ' -
\alpha\, ''- \alpha\, ' \,^{2}
  \right) =  \nonumber \\ \Lambda -  \frac{\kappa}{8
  \pi}(E^2_{(1)}-E^2_{(2)}+B^2_{()}),
\end{eqnarray}
\begin{eqnarray}
E_{(2)}B_{()}=0, \,\,\,
\end{eqnarray}
\begin{eqnarray}
E_{(1)}B_{()}=0, \,\,\,
\end{eqnarray}
and
\begin{eqnarray}
 \label{uno-dos} E_{(1)}E_{(2)}=0,
\end{eqnarray}
where the prime denotes the differentiation $d/dr$.

Now we must consider the Maxwell equations. The contravariant
density components of~(\ref{siete}) are
\begin{eqnarray}
\sqrt{-g} \,  F^{a b}=-2E_{(1)} \, e^{\gamma} \,
\delta^{[a}_{x^0}\, \delta^{b]}_{x^1}-2 E_{(2)} \, e^{\beta} \,
\delta^{[a}_{x^0}\, \delta^{b]}_{x^2} \\ \nonumber +2 B_{()} \,
e^{\alpha} \, \delta^{[a}_{x^1}\, \delta^{b]}_{x^2 }.
\end{eqnarray}
Then, from the source-free Maxwell equations $(\sqrt{- g}\, F^{a
b})_{, b}= 0$ we obtain the constraints
\begin{eqnarray}
\label{ecuacion electrica} E_{(1)}= C_1 \, e^{-\gamma}, \,\,\,
B_{()}= C_2 \, e^{-\alpha},
\end{eqnarray}
and $E_{(2)}$ remains to be an arbitrary function of the
$x^1$-coordinate (here $C_1$ and $C_2$ are constants of
integration).

Although for these (2+1)-dimensional solutions no duality rotation
exists, there is a duality mapping which connects the electric
field $E_{(1)}$ and the scalar magnetic field $B_{()}$.
Effectively, from~(\ref{cero-cero})--(\ref{uno-dos}) we see that
the Einstein-Maxwell equations are invariant under the duality
mapping
\begin{eqnarray}
\label{duality mapping} \alpha \longrightarrow \gamma, \,\,\,
\gamma \longrightarrow \alpha, \nonumber \\ E_{(1)}\,
\longrightarrow i \,  B_{()}, B_{()} \, \longrightarrow i \,
E_{(1)}.
\end{eqnarray}
Notice that from the Einstein-Maxwell equations we conclude that
either $E_{(1)}$, or $E_{(2)}$ or the magnetic field $B_{()}$ must
be zero, i.e. we can not have any superposition of these fields in
any three-dimensional static spacetime.

The solution for $E_{(2)}=B_{()}=0$ may be written in the
form~(\ref{metrica estatica}), where
\begin{eqnarray}
\label{BTZ derivado} e^{2\alpha}=e^{-2\beta}=C-\Lambda
(x^1)^2-q^2_{e}ln (x^1), \,\,\, e^{\gamma}=x^1,
\end{eqnarray}
$E_{(1)}=q_{_{e}}/x^1$ and $C$ is a constant of integration. The
solution for $E_{(1)}=E_{(2)}=0$ may be obtained by using
on~(\ref{metrica estatica}) and on~(\ref{BTZ derivado}) the
duality mapping~(\ref{duality mapping}) in the form~\cite{Cataldo}
\begin{eqnarray}
\label{dual} x^0 \longrightarrow i\,x^2,\,\,\, x^2 \longrightarrow
i\,x^0, \,\,\,q_{_{e}}\longrightarrow i\,q_{_{m}}.
\end{eqnarray}
From these equations we conclude that we have a two-dimensional
duality in the plane $x^0/x^2$.

Then, we have that the magnetostatic solution has the
form~(\ref{metrica estatica}), where
\begin{eqnarray}
\label{HW derivado} e^{\alpha}=x^1,
e^{-2\beta}=e^{2\gamma}=C-\Lambda (x^1)^2+q^2_{m}ln (x^1)
\end{eqnarray}
and $B_{()}=q_{_{m}}/x^1$.

The case for $E_{(1)}=B_{()}=0$ completes the solution of the
(2+1)-static Einstein-Maxwell problem. It is interesting to note
that the Maxwell equations do not fix the form of the $E_{(2)}$
component of the electric field and that this is done by the
Einstein equations. This situation also takes place for the
magnetic field in the Reissner-Nordstr\"{o}m solution in
(3+1)-dimensions.

We now shall find the solution for any static (2+1)-dimensional
spacetime, and for the electric field component which is not
related to the scalar magnetic field by a duality mapping.
From~(\ref{cero-cero}) and~(\ref{uno-uno}) we find
\begin{eqnarray}
\label{degamma} (\gamma^{'} \, e^{\alpha-\beta+\gamma})^{'}=-2
\Lambda \, e^{\alpha+\beta+\gamma}
\end{eqnarray}
and from~(\ref{uno-uno}) and~(\ref{dos-dos}) we find
\begin{eqnarray}\label{dealpha}
(\alpha^{'} \, e^{\alpha-\beta+\gamma})^{'}=-2 \Lambda \,
e^{\alpha+\beta+\gamma}
\end{eqnarray}
This system can be solved exactly if we suppose, without any loss
of generality, that $e^{\beta}=1$. Then we have that
\begin{eqnarray}
((\alpha^{'}+\gamma^{'}) \, e^{\alpha+\gamma})^{'}=- 4 \Lambda \,
e^{\alpha+\gamma},
\end{eqnarray}
from which we obtain the following solutions:
\begin{itemize}
\item For $\Lambda=0$ we find
\end{itemize}
\begin{eqnarray}
e^{\alpha}=(Cx^1+D)^{(1-F)},  \,\,\,
e^{\gamma}=(Cx^1+D)^F,
\end{eqnarray}
where $C$, $D$ and $F$ are integration constants. Re-scaling the
$x^1$-coordinate we finally find
\begin{eqnarray}
\label{lambda=0}
ds^2=(x^1)^{2F}(dx^0)^2-(x^1)^{2F}(dx^1)^2-(x^1)^2(dx^2) ^2,
\end{eqnarray}
and
\begin{eqnarray}
E^2_{(2)}=\frac{8\pi F}{\kappa} \, (x^1)^{-2(F+1)};
\end{eqnarray}
from here we see that $F \geq 0$.
\begin{itemize}
\item For $\Lambda=-1/l^2< 0$ we find
\end{itemize}
\begin{eqnarray}
e^{2\alpha}=e^{-2\tilde{G}I^{^{-}}} \left(C \, sinh(2x^1/l)+D \,
cosh(2x^1/l)\right),
\end{eqnarray}
\begin{eqnarray}
e^{2\gamma}=e^{2\tilde{G}I^{^{-}}} \left(C \, sinh(2x^1/l)+D \,
cosh(2x^1/l)\right),
\end{eqnarray}
where
\begin{eqnarray}
I^{^{-}}=\frac{l}{\sqrt{D^2-C^2}} \, \arctan\left(\frac{D
\tanh(x^1/l)+C}{\sqrt{D^2-C^2}} \right).
\end{eqnarray}
\begin{itemize}
\item For $\Lambda=1/l^2> 0$ we find
\end{itemize}
\begin{eqnarray}
e^{2\alpha}=e^{-2\tilde{G}I^{^{+}}} \left(C \, sin(2x^1/l)+D \,
cos(2x^1/l)\right),
\end{eqnarray}
\begin{eqnarray}
e^{2\gamma}=e^{2\tilde{G}I^{^{+}}} \left(C \, sin(2x^1/l)+D \,
cos(2x^1/l)\right),
\end{eqnarray}
where now
\begin{eqnarray}
I^{^{+}}=\frac{l}{\sqrt{D^2+C^2}} \, \arctan\left(\frac{D
\tanh(x^1/l)-C}{\sqrt{D^2+C^2}} \right).
\end{eqnarray}
Here $\tilde{G}$ is a constant of integration and we have
introduced the parameter $l$ related to $\Lambda$ for further use.

From now on we shall assume the spacetime to be statically and
rotationally symmetric, implying the existence of a timelike
Killing vector $\eta=\partial_t$ and a spacelike Killing vector
$\zeta=\partial_{ \varphi}$. This allows us to identify $x^{0}=t$,
$x^1=r$ (the radial coordinate) and $x^2=\varphi$ where $-\infty
<t<\infty$, $0<r<\infty$ and $0<\varphi<2\pi$. In this case
$E_r(r)$ is the radial electric field, $E_{\varphi}(r)$ is the
azimuthal electric field and $B(r)$ is the scalar magnetic field.

The rotationally symmetric spacetimes may be classified as
follows: For the radial electric field we have the BTZ
solution~(\ref{BTZ}), which is obtained from~(\ref{BTZ derivado})
by identifying $C=-M$ and $\Lambda=-1/l^2$. For the scalar
magnetic field we have the HW solution~(\ref{HWS}), which is
obtained from~(\ref{HW derivado}) by making the same
identifications for $C$ and $\Lambda$, and re-scaling the
$r$-coordinate.

The case for the azimuthal electric field has not been studied up
to now. For this case we shall consider a particular solution
given by $\tilde{G}=0$. In this case the obtained solutions take a
simple form. Effectively, making $\rho^{2}=C \, \sinh(2r/l)+D \,
\cosh(2r/l)$ and $\rho^{2}=C \, \sin(2r/l)+D \, \cos(2r/l)$ we
have~\begin{eqnarray} ds^2=\rho^2dt^2-\frac{d\rho^2}{q^2/\rho^2
\pm \rho^2/l^2}-\rho^2d\varphi^2,
\end{eqnarray}~for $\Lambda=-1/l^2$ (the plus sign) and $\Lambda=1/l^2$
(the minus sign) respectively. In this case the azimuthal
component of the electric field is given by
$E_{(2)}=\frac{q}{\sqrt{G} \, r^2}$. If $\Lambda=0$, we have a
particular metric of~(\ref{lambda=0}) specified by $F=1$. On the
other hand, if $q=0$ and $\Lambda=-1/l^2<0$, the vacuum state is
obtained, namely, what is to be regarded as empty anti-de Sitter
space.

It is interesting to note that these solutions are related to a
subset of solutions to the gravitating $O(3)$ $\sigma$ model in
(2+1)-dimensions~\cite{Clement2}.

Let us now study the kind of solutions induced by~(\ref{2+1 tensor
de Maxwell}) (with $B_{()}=:B_{(3)}$) on the time-independent
cylindrically symmetric spacetimes when the Einstein-Maxwell
theory is considered. Any cylindrically static spacetime admits
the Killing vectors $\eta =\partial _{t,}$ $\zeta =\partial
_{\varphi ,\text{ }}\xi =\partial _{z}$ and the metric can be
written in the form
\begin{eqnarray}
\label{metrica cilindrica} ds^{2}=e^{2\alpha (r)}dt^{2}-e^{2\beta
(r)}\ dr^{2}-e^{2\gamma (r)}d\varphi^2
-e^{2\delta (r)}dz^{2}.  
\end{eqnarray}
We now introduce an orthonormal 1-form basis given by
\begin{eqnarray}
\label{4tetrada} \theta^{(0)}=e^{\alpha(r)} dt, \,\,\,
\theta^{(1)}=e^{\beta(r)} dr, \nonumber
\\ \theta^{(2)}=e^{\gamma(r)} d\varphi, \,\,\,
\theta^{(3)}=e^{\delta(r)}dz.
\end{eqnarray}
In this case we shall take the Maxwell tensor in the
form
\begin{eqnarray}
\label{F cilindrico} F=-E_{(1)} \theta^{(0)}\wedge
\theta^{(1)}-E_{(2)} \theta^{(0)} \wedge \theta^{(2)}  \nonumber
\\ + B_{(3)} \theta^{(1)} \wedge \theta^{(2)},
\,\,\,\,\,\,\,\,\,\,\,\,\,\,\,\,\,
\end{eqnarray}
Now $E_{(1)}$ is a radial electric field, $E_{(2)}$ is an
azimuthal electric field and $B_{(3)}$ is a magnetic field along
the z direction (we can include here for example an electric
component $E_{(3)}$ pointing along the $z$ direction but this
component is related by a duality rotation to $B_{(3)}$). These
components of the electric and magnetic field are measured by a
locally lorentzian observer in the orthonormal
basis~(\ref{4tetrada}). From~(\ref{tensor electromagnetico}) we
obtain for~(\ref{F cilindrico})
\begin{eqnarray}
T^{3+1}= -\frac{1}{8 \pi}
\left[-(E^2_{(1)}+E^2_{(2)}+B^2_{(3)})\theta^{(0)}\otimes
\theta^{(0)} \right. \nonumber \\
+(E^2_{(1)}-E^2_{(2)}-B^2_{(3)})\theta^{(1)}\otimes \theta^{(1)}
\nonumber \\ -(E^2_{(1)}-E^2_{(2)}+B^2_{(3)})\theta^{(2)}\otimes
\theta^{(2)} \\ \nonumber
+(B^2_{(3)}-E^2_{(1)}-E^2_{(2)})\theta^{(3)}\otimes \theta^{(3)}
\\ \nonumber
 +2 E_{(2)} B_{(3)}(\theta^{(0)}\otimes \theta^{(1)}+\theta^{(1)}
 \otimes \theta^{(0)})
\\
\nonumber -2 E_{(1)} B_{(3)}(\theta^{(0)}\otimes
\theta^{(2)}+\theta^{(2)}\otimes \theta^{(0)})
\\ \nonumber
\left. +2 E_{(1)} E_{(2)}(\theta^{(1)}\otimes
\theta^{(2)}+\theta^{(2)}\otimes \theta^{(1)})
 \right].
\end{eqnarray}
The Einstein equations with cosmological constant are
\begin{eqnarray}
\label{4cero-cero} e^{- 2 \beta}\,\left(-\gamma\, ''- \gamma\, '
\,^{2} + \beta\, \, '\gamma\, ' -\delta\, ''-\delta\,
'^{2}+\beta\, '\, \delta ' \right. \\ \nonumber \left. -\gamma\, '
\, \delta\, ' \right) = \frac{\kappa}{8  \pi} \,
(E^2_{(1)}+E^2_{(2)}+B^2_{(3)})+ \Lambda
\end{eqnarray}
\begin{eqnarray}
\label{4uno-uno} e^{- 2\, \beta}(- \alpha\, ' \, \gamma\, '
-\alpha\, '\delta\, '-\gamma\, '\delta\, ') \nonumber \\=
\frac{\kappa}{8 \pi} (E^2_{(1)}-E^2_{(2)}-B^2_{(3)})+ \Lambda
\end{eqnarray}
\begin{eqnarray}
\label{4dos-dos}  e^{- 2 \beta}\,\left(-\alpha\, ''- \alpha\, '
\,^{2} + \alpha\, \, '\beta\, ' -\delta\, ''-\delta\,
'^{2}-\alpha\, '\, \delta '\right. \\ \nonumber \left. +\beta\, '
\, \delta\, ' \right)   = - \frac{\kappa}{8
  \pi}(E^2_{(1)}-E^2_{(2)}+B^2_{(3)})+ \Lambda,
\end{eqnarray}
\begin{eqnarray}
\label{4tres-tres}  e^{- 2 \beta}\,\left(-\alpha\, ''- \alpha\, '
\,^{2} + \alpha\, \, '\beta\, ' -\gamma\, ''-\gamma\,
'^{2}-\alpha\, '\, \gamma '\right. \\ \nonumber \left. +\beta\, '
\, \gamma\, ' \right) = \frac{\kappa}{8
  \pi}(B^2_{(3)}-E^2_{(1)}-E^2_{(2)})+ \Lambda,
\end{eqnarray}
and
\begin{eqnarray}
\label{4cero-uno} E_{(2)}B_{(3)}=0,
 \,\,\, E_{(1)}B_{(3)}=0, \,\,\, E_{(1)}E_{(2)}=0,
\end{eqnarray}
where the prime denotes the differentiation $d/dr$. The
source-free Maxwell equations are now
\begin{eqnarray}
\label{4ecuacion electrica} E_{(1)}= \tilde{C}_1 \,
e^{-\gamma-\delta}, \,\,\,  B_{(3)}= \tilde{C}_2 \,
e^{-\alpha-\delta},
\end{eqnarray}
and $E_{(2)}$ remains to be an arbitrary function of the
r-coordinate (here $\tilde{C}_1$ and $\tilde{C}_2$ are constants
of integration).

From eqs.~(\ref{4cero-cero})-(\ref{4ecuacion electrica}) we see
that, under the following duality mappings, the Einstein-Maxwell
equations are invariant:
\begin{eqnarray}
\label{alpha gamma} \alpha \longrightarrow \gamma, \gamma
\longrightarrow \alpha, E_{(1)} \longrightarrow  i \, B_{(3)},
B_{(3)} \longrightarrow i \, E_{(1)},
\end{eqnarray}
\begin{eqnarray}
\label{alpha delta} \alpha \longrightarrow \delta, \delta
\longrightarrow \alpha, E_{(1)} \longrightarrow  i \, E_{(2)},
E_{(2)} \longrightarrow i \, E_{(1)},
\end{eqnarray}
\begin{eqnarray}
\label{gamma delta} \gamma \longrightarrow \delta, \delta
\longrightarrow \gamma, E_{(2)} \longrightarrow   B_{(3)}, B_{(3)}
\longrightarrow E_{(2)}.
\end{eqnarray}

We conclude from the cylindrical Einstein-Maxwell equations that
we can not consider simultaneously all components for the
electromagnetic field given by~(\ref{F cilindrico}), and thus we
have that either the radial, or the azimuthal electric field, or
the magnetic field must be zero.

Thus all the considered solutions may be obtained from just one
given solution. Witten~\cite{Witten} obtained the solution for a
magnetic field along the $z$-axis for $\Lambda=0$. In this case
the metric has the form
\begin{eqnarray}
ds^2=r^{2m^2}\, G^2 (dt^2-dr^2)-G^{-2}d\varphi^2-r^2G^2dz^2,
\end{eqnarray}
where $G=Ar^m+Br^{-m}$ with $A$, $B$ and $m$ constants. We shall
consider in detail one particular solution for this metric: the
well known Melvin-Bonnor universe, which is given by~\cite{Melvin}
\begin{eqnarray}
ds^2=\left(1+\frac{b^2}{4}r^2 \right)^2 \left
(dt^2-dr^2-dz^2\right ) \nonumber \\ - r^2
\left(1+\frac{b^2}{4}r^2 \right)^{-2} d\varphi^2,
\end{eqnarray}
and $B_{(3)}=(b/\sqrt{G} )(1+\frac{b^2}{4}r^2)^{-2}$. This metric
may be generalized to the case where $\Lambda \neq 0$. In another
gauge it takes the form
\begin{eqnarray}
\label{AdS Soliton}
ds^2=r^2(dt^2-dz^2)-\left(\frac{C}{r}-\frac{q^2_m}{r^2}-
\frac{\Lambda}{3} r^2\right)^{-1}dr^2 \nonumber \\
-\left(\frac{C}{r}-\frac{q^2_m}{r^2}- \frac{\Lambda}{3} r^2
\right)d\varphi^2,
\,\,\,\,\,\,\,\,\,\,\,\,\,\,\,\,\,\,\,\,\,\,\,\,\,\,\,
\end{eqnarray}
with $C$ a constant, $q_m$ a magnetic charge and
$B_{(3)}=q_m/(\sqrt{G}\, r^2)$. If we apply the corresponding
duality mapping~(\ref{alpha gamma}), we obtain the electric
Melvin-Bonnor analog in the form
\begin{eqnarray}\label{MB radial}
ds^2=\left(\frac{C}{r}+\frac{q^2_e}{r^2}- \frac{\Lambda}{3}
r^2\right)dt^2 \,\,\,\,\,\,\,\,\,\,\,\,\,\,\,\, \nonumber \\ -
\left(\frac{C}{r}+\frac{q^2_e}{r^2}- \frac{\Lambda}{3} r^2
\right)^{-1}dr^2 -r^2(d\varphi^2-dz^2),
\end{eqnarray}
and $E_{(1)}=q_e/(\sqrt{G}r^2)$.

Notice that in~\cite{Melvin} the authors did not consider~(\ref{MB
radial}). They only considered the Melvin-Bonnor universe with a
purely electric or a purely magnetic field pointing along the $z$
direction, which are connected by a duality rotation. We must note
that eq.~(\ref{MB radial}) coincides with the solution (2.3) given
in~\cite{Lemos}. The magnetic counterpart of the solution~(\ref{MB
radial}) in AdS background is given in~\cite{Lemos2}. Here the
range of the coordinates dictates the topology of this solution.
We can have cylindrical, planar or toroidal topology. For
instance, a cylindrical topology has the range $0<r<\infty$,
$0<\varphi< 2 \pi$, and $-\infty<z<\infty$.

It is interesting to note also, that the metric~(\ref{AdS
Soliton}) is related to the 4-dimensional AdS
soliton~\cite{Horowitz} with an electric field (in this case $C=0$
and $\Lambda=-3/l^2$). The magnetic soliton is connected to the
electric one by a duality rotation. The corresponding AdS black
hole may be obtained from~(\ref{AdS Soliton}) (with $C=0$ and
$\Lambda=-3/l^2$) by the duality mapping~(\ref{gamma delta}) and
after one must analytically continue the obtained metric, $t
\longrightarrow i z$ and $z \longrightarrow i t$.

In conclusion, in a static rotationally symmetric
(2+1)-dimensional gravity only the electrostatic spacetime with a
radial electric field is a black-hole solution; the
self-consistent gravitational fields with azimuthal electric or
scalar magnetic field are not black-holes. In this case we also
can not have a superposition of a radial electric field with an
azimuthal electric nor with a magnetic field. Then we have the
situation analogous to (3+1)-dimensional cylindrical static
gravity considered above, where any superposition of the
electromagnetic components given by~(\ref{F cilindrico}) is not
suitable. The above situation is induced by the Maxwell
tensor~(\ref{2+1 tensor de Maxwell}) (with $B_{()}=:B_{(3)}$) in
Einstein-Maxwell theory for cylindrically symmetric spacetimes. On
the other hand, the magnetic Melvin-Bonnor universe (as well as
the solution with an azimuthal electric field) is not a black
hole; while the electric Melvin-Bonnor universe~(\ref{MB radial})
with a radial electric field and $\Lambda=0$ has a cosmological
horizon as well as the electrovacuum (2+1)-static solution found
in~\cite{Gott}, which is the same solution~(\ref{BTZ}) for
$\Lambda=0$. From~(\ref{MB radial}) we see that if $\Lambda<0$ we
have a cylindrical charged black hole or black
string~\cite{Lemos}, for which the BTZ solution~(\ref{BTZ}) is the
(2+1)-dimensional absolute analog. Furthermore, if $q=0$, we have
a black string solution only because we have included a negative
cosmological constant and then this spacetime is asymptotically
anti-de Sitter as well as occurs in (2+1)-dimensions.

I thank Paul Minning and Carol Mu\~noz for carefully reading and
typing this manuscript respectively. I also wish to thank Alberto
Garc\'\i a for informative comments. This work was supported by
CONICYT through Grant FONDECYT N$^0$ 1010485 and by Direcci\'on de
Promoci\'on y Desarrollo de la Universidad del B\'\i o-B\'\i o.

\end{document}